\documentclass[aps,prl,floatfix,twocolumn,reprint,amsmath,amssymb,superscriptaddress]{revtex4-1}

\usepackage{amsfonts}
\usepackage{mathrsfs}
\usepackage{amsmath}
\usepackage{color}
\usepackage{graphicx}
\usepackage{bm}
\usepackage{amssymb}
\usepackage{xspace}
\usepackage{epstopdf}
\usepackage{dcolumn}
\usepackage{longtable}
\usepackage{multirow}
\usepackage{float}
\usepackage{comment}
\usepackage{soul}

\usepackage[colorlinks=true, letterpaper=true, pdfstartview=FitV, linkcolor=blue, citecolor=blue, urlcolor=blue]{hyperref}

\makeatother

\begin{document}

\title{Anomalous Hall Effect in Type IV 2D Collinear Magnets}

\author{Ling Bai}
\affiliation{Key Lab of Advanced Optoelectronic Quantum Architecture and Measurement (MOE), Beijing Key Lab of Nanophotonics and Ultrafine Optoelectronic Systems, and School of Physics,Beijing Institute of Technology, Beijing 100081, China}
\affiliation{International Center for Quantum Materials, Beijing Institute of Technology, Zhuhai, 519000, China}

\author{Run-Wu Zhang}
\affiliation{Key Lab of Advanced Optoelectronic Quantum Architecture and Measurement (MOE), Beijing Key Lab of Nanophotonics and Ultrafine Optoelectronic Systems, and School of Physics,Beijing Institute of Technology, Beijing 100081, China}
\affiliation{International Center for Quantum Materials, Beijing Institute of Technology, Zhuhai, 519000, China}

\author{Wanxiang Feng}
\email{wxfeng@bit.edu.cn}
\affiliation{Key Lab of Advanced Optoelectronic Quantum Architecture and Measurement (MOE), Beijing Key Lab of Nanophotonics and Ultrafine Optoelectronic Systems, and School of Physics,Beijing Institute of Technology, Beijing 100081, China}
\affiliation{International Center for Quantum Materials, Beijing Institute of Technology, Zhuhai, 519000, China}

\author{Yugui Yao}
\email{ygyao@bit.edu.cn}
\affiliation{Key Lab of Advanced Optoelectronic Quantum Architecture and Measurement (MOE), Beijing Key Lab of Nanophotonics and Ultrafine Optoelectronic Systems, and School of Physics,Beijing Institute of Technology, Beijing 100081, China}
\affiliation{International Center for Quantum Materials, Beijing Institute of Technology, Zhuhai, 519000, China}

\date{\today}

\begin{abstract}
We identify a previously unrecognized class of two-dimensional (2D) collinear magnetic phase that extends beyond the established categories of ferromagnets, antiferromagnets, and altermagnets. These type IV 2D collinear magnets exhibit spin-degenerate bands in the nonrelativistic limit, yet support time-reversal symmetry-breaking responses, such as the anomalous Hall effect (AHE), despite having zero net magnetization. Based on spin layer group analysis, we derive the symmetry criteria for this phase and perform first-principles calculations to screen viable candidate materials from 2D databases. Using monolayer Hf$_2$S as a prototype, we demonstrate that in the absence of spin-orbit coupling, the bands are spin degenerate, while its inclusion induce an AHE driven by spin-polarized and even spin-neutral currents, accompanied by a symmetry-protected, truly full-space persistent spin texture. These findings expand the classification of magnetic phases and broaden avenues for realizing unconventional spintronic functionalities in two dimensions.

\end{abstract}

\maketitle

Magnetism, a time-honored yet ever-evolving domain in solid-state physics, encompasses a rich spectrum of physical phenomena and enduring challenges. In magnetic solids, exchange interactions between localized magnetic moments govern the parallel or antiparallel spin alignments, which historically define two archetypal forms of magnetic order: ferromagnets (FMs) and antiferromagnets (AFMs). AFMs, characterized by the absence of stray fields and ultrafast terahertz spin dynamics, have emerged as compelling candidates for ultradense and high-speed spintronic applications, potentially outperforming traditional FM-based systems. On account of this, antiferromagnetic spintronics~\cite{Baltz2018,Zelezny2018,Jungwirth2018,Smejkal2018} has attracted considerable attention in the past decade. Nevertheless, the lack of macroscopic magnetization in AFMs presents notable hurdles, particularly in realizing efficient electrical readout and manipulation of binary ‘0’ and ‘1’ states via N\'eel order switching. This inherent limitation has prompted the pursuit of unconventional magnetic phases, among which altermagnets (AMs)~\cite{SmejkalPRX2022a,SmejkalPRX2022b} have recently been proposed as a third magnetic archetype, unifying the functional advantages of both FMs and AFMs. Although devoid of net magnetization, AMs exhibit a suite of ferromagnetic-like phenomena, including nonrelativistic band spin splitting and the anomalous Hall effect (AHE)~\cite{Smejkal2020,ZX-Feng2022}, offering promising mechanisms for electrical signal control.

For practical device applications, two-dimensional (2D) magnetic systems are particularly advantageous due to their exceptional scalability, greater tunability, and superior integration compatibility compared to bulk materials. Following the breakthroughs in achieving stable long-range magnetic ordering in mono- and few-layer Cr$_2$Ge$_2$Te$_6$~\cite{GongNature2017} and CrI$_3$~\cite{B-Huang2017}, a rapidly growing library of atomically thin 2D FMs and AFMs has been established through both chemical synthesis and mechanical exfoliation from bulk crystals~\cite{QH-Wang2022,Kajale2024,BY-Zhang2024}. In contrast, the development of 2D AMs remains in its infancy. To date, only a limited number—approximately two dozen—of such candidates have been theoretically predicted, in stark contrast to over 200 identified three-dimensional AMs~\cite{BaiAFM2024}, largely due to the more restrictive symmetry requirements in reduced dimensions. Moreover, the experimental realization of 2D altermagnetism remains elusive. This highlights a pressing need to identify new classes of unconventional 2D magnets, offering broader material choices and improved experimental feasibility for future high-performance spintronic devices.

\begin{table*}
	\renewcommand{\arraystretch}{1.5}
	\caption{Nontrivial spin Laue groups and representative material candidates for type IV 2D collinear magnets.  The first column groups crystallographic layer groups sharing the same Laue group ($\mathbf{G}$). The second column lists symmetry operations in the halving subgroup $\mathbf{H}$ of $\mathbf{G}$. The other coset $\mathbf{G-H}$, an essential component of a nontrivial spin Laue group,  contains symmetry operations connecting opposite-spin sublattice.}
	\label{tab:Tab1}
	\begin{ruledtabular}
		\begin{tabular}{ccc}
			Layer groups (Laue groups $\mathbf{G}$) & $\mathbf{H}$ (Laue group) & Material candidates \\
			\hline
			3-7 ($C_{2h}$)           &  $\{E, \bar{E}\} (C_i)$  & MnSH$_2$O$_4$         \\ 
			19-48 ($D_{2h}$)       &  $\{E, \bar{E},C_{2x},M_{x}\}$ or  $\{E, \bar{E},C_{2y},M_{y}\}(C_{2h})$ &  MnSbS$_2$Cl, Co$_2$P$_4$O$_8$ \\
			73-75 ($C_{6h}$)      &  $\{E, \bar{E},C_{3z}^{+},C_{3z}^{-}, \bar{C}_{3z}^{+},  \bar{C}_{3z}^{-}\} (C_{3i})$ & bilayer FeHfBr$_6$ \\  
			76-80 ($D_{6h}$)     &  $\{E, \bar{E},C_{3z}^{+},C_{3z}^{-}, \bar{C}_{3z}^{+},  \bar{C}_{3z}^{-}, C_{2[100]}, C_{2[010]}, C_{2[110]}, M_{[100]},M_{[010]},M_{[110]} \}(D_{3d})$  & BaMn$_2$S$_3$, Hf$_2$S  \\
		\end{tabular}
	\end{ruledtabular}
\end{table*}

In this Letter, leveraging spin layer group analysis, we identify a previously overlooked class of type IV 2D collinear magnets, beyond the known FMs, AFMs, and AMs. This new phase exhibits nonrelativistic spin-degenerate band structures, akin to conventional AFMs, yet simultaneously supports time-reversal symmetry-breaking responses, such as the AHE—typically considered a hallmark of FMs and AMs. Through first-principles calculations, we screen viable material candidates from comprehensive 2D material databases, covering all symmetry-allowed nontrivial spin red groups. Taking monolayer Hf$_2$S as a representative example, we demonstrate the emergence of nonrelativistic spin-degenerate bands, which evolve into distinct anomalous transport behavior upon inclusion of spin-orbit coupling (SOC). Notably, we reveal an AHE driven by spin-neutral currents, accompanied by a symmetry-protected, truly full-space persistent spin texture (TFPST). Our work not only enriches the conceptual framework of low-dimensional magnetism but also provides a robust platform for engineering next-generation spintronic functionalities through this newly uncovered class of 2D magnetic materials.

We begin by classifying 2D collinear magnets through spin group analysis. In general, a spin group can be regarded as a direct product of a spin-only group ($\mathbf{r}_{s}$) and a nontrivial spin group ($\mathbf{R}_{s}$)~\cite{Litvin1974,Litvin1977}. For collinear spin configurations, the spin-only group is given by $\mathbf{r}_{s} = \mathbf{C_{\infty}} + \bar{C}_{2} \mathbf{C_{\infty}}$, where $ \mathbf{C_{\infty}}$ denotes arbitrary spin rotations around the common spin axis, and $\bar{C}_{2}$ represents a twofold rotation perpendicular to this axis ($C_{2}$), accompanied by inversion in spin space, which is actually realized by the time-reversal operation ($T$)~\cite{Andreev1980}.  $\mathbf{R}_{s}$ contains the elements $ [R_i || R_j]$, where $R_i$ and $R_j$ are symmetry operations acting independently in spin space and crystal space, respectively. In collinear magnets, the allowed spin-space operations are $R_i \in \{E\}$ or $\{E, C_2\}$, where $E$ denotes the identity operation. The crystal-space operation $R_j$ is constrained within the crystallographic Laue group ($\mathbf{G}$) that is derived from 2D layer groups. This is because that inversion symmetry ($\bar{E}$) is always effectively preserved in the nonrelativistic band structures of collinear magnets, regardless of whether the crystals possess $\bar{E}$ or not~\cite{note1}.

According to the distinct nontrivial spin Laue groups, 2D collinear magnets can be classified into four categories. Type I, denoted as $\mathbf{R}_{s}^\textnormal{I} = [E || \mathbf{G}]$, contains only crystal-space transformations and does not enforce spin degeneracy at any momentum point in the 2D Brillouin zone (BZ), thereby corresponding to FMs (or ferrimagnets).  Type II is represented by $\mathbf{R}_{s}^\textnormal{II} = [E || \mathbf{G}]+[C_{2}||\mathbf{G}]$, which describes conventional collinear AFMs with zero net magnetization. In this case, operations such as $[C_{2}||E]$ or $[C_{2}|| \bar{E}]$ lead to spin degeneracy throughout the entire BZ. It is apparent that $[C_{2}||E]\epsilon(s,k)=\epsilon(-s,k)$ and $\bar{C}_{2}[C_{2}|| \bar{E}]\epsilon(s,k)=\epsilon(-s,k)$, where $s$ and $k$ represent the spin and crystal momentum indices of the electronic bands, respectively~\cite{note2,XB-Chen2025}.  It is noted that the spin-only operation $\bar{C}_{2}$ plays an auxiliary role in the latter case.  Type III, corresponding to AMs, is defined as $\mathbf{R}_{s}^\textnormal{III} = [E || \mathbf{H}]+[C_{2}||\mathbf{G-H}]=[E || \mathbf{H}]+[C_{2}||A\mathbf{H}]$, where $\mathbf{H}$ denotes a halving subgroup of $\mathbf{G}$, and $A$, belonging to $\mathbf{G-H}$, represents a real-space rotation. In contrast to the three-dimensional case, where the condition is $A\notin \{\tau, \bar{E}\}$, the 2D scenario imposes a more stringent constraint, requiring $A\notin \{\tau, \bar{E}, C_{2z}, M_{z}\}$ (assuming the 2D collinear magnet lies in the $x$-$y$ plane). This restriction arises from the in-plane nature of the crystal momentum $k$, as well as the fact that the following symmetry operations,
\begin{eqnarray}
	\bar{C}_{2}\left[C_{2}|| C_{2z}\right]\epsilon(s,k)&=&\epsilon(-s,k), \label{eq:C2z} \\
	\left[C_{2}||M_{z}\right]\epsilon(s,k)&=&\epsilon(-s,k), \label{eq:Mz}
\end{eqnarray}
ensure spin degeneracy throughout the 2D BZ, and therefore, prevent the emergence of alternating spin splitting. This underlying principle has been applied in the recent theoretical identification of 2D AMs~\cite{ZengPRB2024a,PanPRL2024,LiuPRL2024,PanPRL2024,ZengPRB2024b}.

The remaining type IV, proposed in the present work, is defined as
\begin{eqnarray}
	\mathbf{R}_{s}^\textnormal{IV} = [E || \mathbf{H}] + [C_{2} || \mathbf{G-H}] = [E || \mathbf{H}] + [C_{2} || B\mathbf{H}], \label{eq:t4}
\end{eqnarray}
with the requirements $B \notin \{\tau, \bar{E}\}$ and $\{C_{2z},M_{z}\} \subseteq B\mathbf{H}$. As evident from Eqs.~\eqref{eq:C2z}--\eqref{eq:Mz}, the presence of $[C_{2}|| C_{2z}]$ or $[C_{2}||M_{z}]$ operations enforces spin degeneracy across the 2D BZ.  By filtering the Laue groups associated with the corresponding layer groups (see supplemental Table~\textcolor{blue}{S1}~\cite{SuppMater}), we have identified four nontrivial spin Laue groups that fall under the type IV classification, as summarized in Table~\ref{tab:Tab1}. This newly proposed magnetic phase is confined to monoclinic, orthorhombic, and hexagonal crystal systems, in contrast to type III, which can also occur in tetragonal systems. Table~\ref{tab:Tab1} lists representative material candidates identified from the Computational 2D Materials Database (C2DB)~\cite{C2DB1,C2DB2}, the Materials Cloud two-dimensional crystals database (MC2D)~\cite{MC2D}, and the van der Waals Bilayer Database (BiDB)~\cite{BiDB}.  Interestingly, we find that the type IV classification can arise from the stacking of two magnetic monolayers, offering a practical and experimentally accessible route toward its realization in 2D heterostructures.  The crystal and band structures of these material candidates are shown in supplemental Table~\textcolor{blue}{S2} and Figs.~\textcolor{blue}{S1} and~\textcolor{blue}{S2}~\cite{SuppMater}.

\begin{figure*}
	\centering
	\includegraphics[width=2\columnwidth]{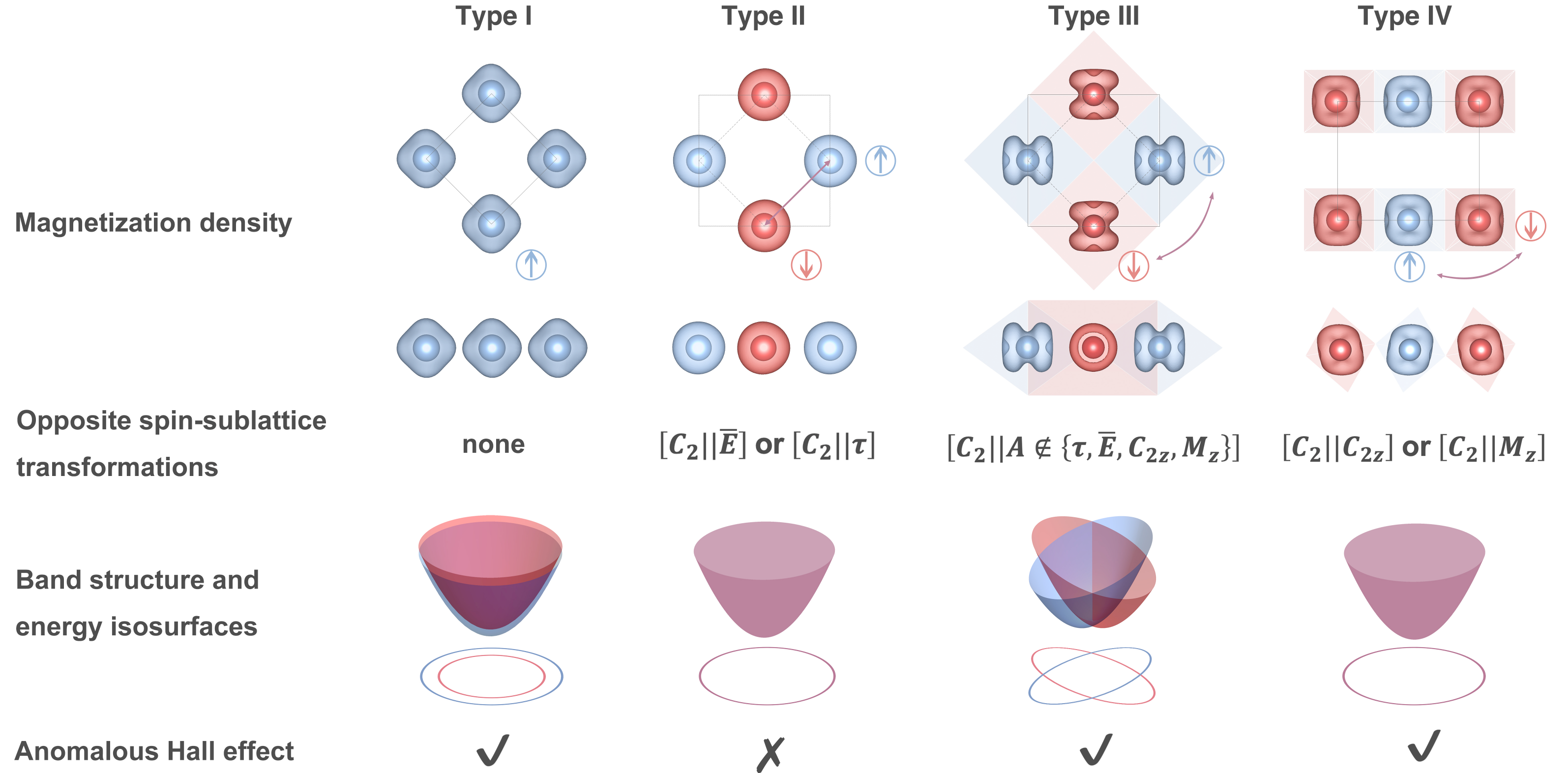}
	\caption{Illustration (in columns) of the four phases of 2D collinear magnets. The first row presents models showing the top and side views of collinear spin arrangements and their magnetization density isosurfaces, with opposite spin directions depicted in blue and salmon. The second row illustrates the opposite-spin-sublattice transformations characteristic of each phase, corresponding to the purple arrows in the first row. The third row displays schematic representations of band structures and their corresponding energy isosurfaces. The last row summarizes the presence or absence of the AHE.}
	\label{fig1}
\end{figure*}

Despite spin degeneracy in the nonrelativistic limit, the type IV phase enables striking $T$-breaking responses, such as the AHE, when SOC is taken into account. As a simplest example, starting from the layer group $p11m$ (No. 4), the derived Laue group $\mathbf{G}$ is $C_{2h}$; then, using the halving subgroup $\mathbf{H} = C_i$ to constructing the nontrivial spin Laue group, written by $\mathbf{R}_{s} = [E||\{E,\bar{E}\}]+[C_{2}||\{C_{2z},M_{z}\}]$, which ensures spin degeneracy throughout the 2D BZ. Upon including SOC and lying the N\'eel vector in the $x$-$y$ plane, the magnetic layer group is $p11m$ (4.1.12), where the symmetry $M_z$ permits a nonzero anomalous Hall conductivity (AHC) $\sigma_{xy}$.  This finding challenges the previously held notion that nonrelativistic band exchange splitting is a prerequisite for the emergence of the AHE~\cite{NagaosaRMP2010}. 

The characteristics of the type IV phase positively set it apart from type I–III phases. In type II, the AHE remains absent even including SOC, while in types I and III, spin degeneracy is lifted in the absence of SOC. Although spin degeneracy and the AHE can appear simultaneously in noncoplanar AFMs~\cite{ShindouPRL2001,FengNC2020,XD-Zhou2023}, the AHE in those systems does not require SOC and originates from a real-space Berry phase effect, situating it within the realm of topological Hall physics. Noncoplanar spin textures are inherently more complex to detect and manipulate compared to collinear configurations, and noncoplanar AFMs are beyond the scope of this study.  Therefore, we identify the type IV phase as a distinct class of 2D collinear magnets, with its relationship to the other three types illustrated in Fig.~\ref{fig1}.

\begin{figure}
	\centering
	\includegraphics[width=1\columnwidth]{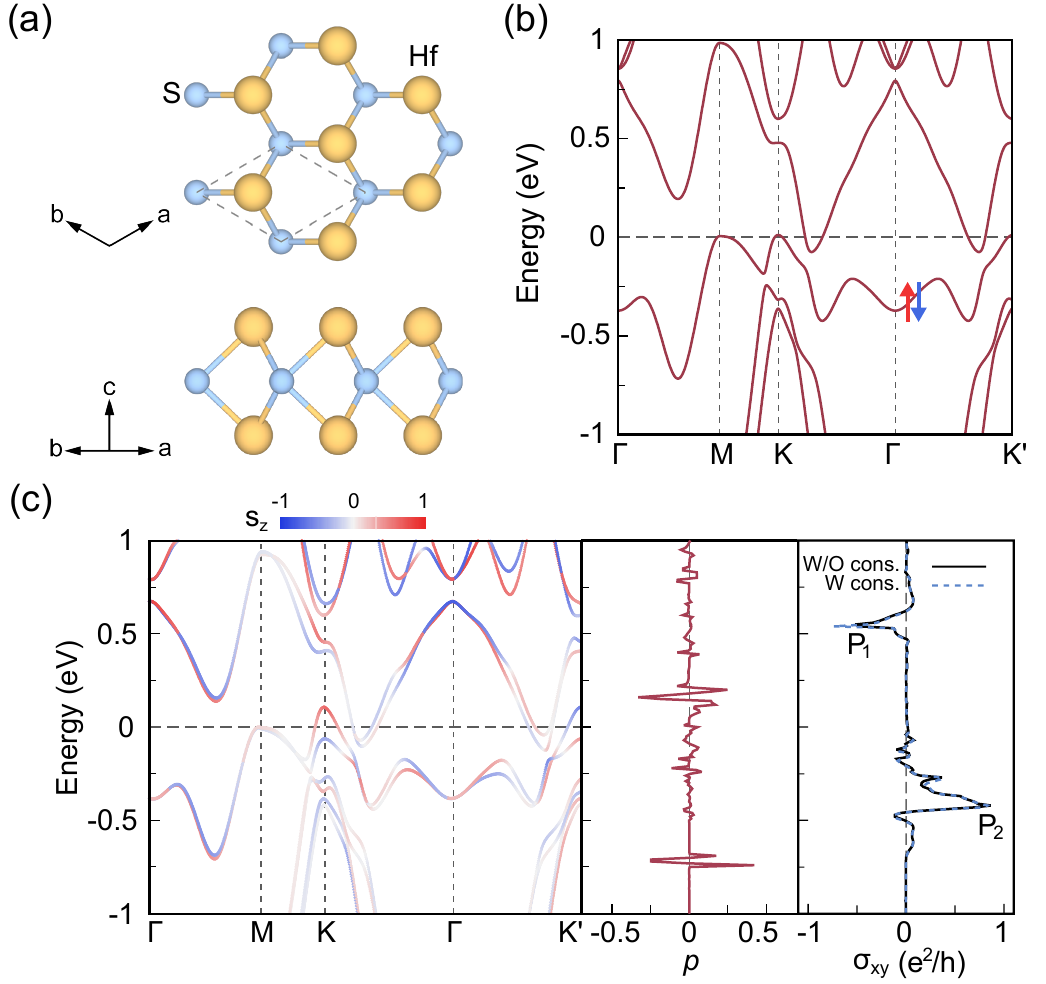}
	\caption{(a) Top and side views of the crystal structure of monolayer Hf$_{2}$S. (b) Nonrelativistic band structure. (c) $s_z$-resolved relativistic band structure (left panel), spin polarizability $p$ (middle panel), and anomalous Hall conductivity $\sigma_{xy}$ (right panel), with the N\'eel vector oriented along the [120] direction ($y$-axis).  The result of $\sigma_{xy}$ with rigidly constrained zero net magnetization is shown for comparison.}
	\label{fig2}
\end{figure}

Next, we take a realistic material, monolayer Hf$_{2}$S, to portray the key features of the type IV magnetic phase. Hf$_{2}$S is a 2D van der Waals material with an anti-TaS$_2$-type crystal structure, known as an electride with excellent chemical stability~\cite{KangSA2020}. Monolayer Hf$_{2}$S belongs to the layer group $p\bar{6}m2$ (No. 78) [Fig.~\ref{fig2}(a)] and exhibits collinear antiparallel magnetic ordering~\cite{LiuPRB2022,ZhangJPCC2023}. The spin layer group of monolayer Hf$_{2}$S includes the following symmetry operations:
\begin{equation}
	\begin{aligned}
		&[E || \{E,C_{3z}^{+},C_{3z}^{-}, M_{[100]}, M_{[010]},M_{[110]}\}]  \\
		&+[C_{2}||\{\bar{C}_{6z}^{+},\bar{C}_{6z}^{-},C_{2[1\bar{1}0]},C_{2[120]}, C_{2[210]},M_{z}\}],
	\end{aligned}
\end{equation}
which corresponds to the last nontrivial spin Laue group listed in Table~\ref{tab:Tab1}. The symmetry operation $[C_{2}||M_{z}]$ enforces nonrelativistic spin degeneracy across the entire 2D BZ, as illustrated in Fig.~\ref{fig2}(b).  When the N\'eel vector is aligned along the [120] direction ($y$-axis), the magnetic layer group becomes $cm^\prime 2^\prime m$ (35.5.216), comprising four symmetry operations: $\{E, M_{z}, C_{2[120]}T, M_{[100]}T\}$. The relativistic band structure, shown in Fig.~\ref{fig2}(c), exhibits full spin splitting across the BZ.  Although the bands at certain points, such as the $\Gamma$ point, appear to be accidentally degenerate, they actually correspond to one-dimensional irreducible co-representations~\cite{ZY-Zhang2023}.  Since $M_{z}$ preserves a nonzero $z$-component of spin but suppresses the $x$- and $y$-components, i.e., $s_{z} \neq 0$ and $s_{x,y} = 0$ (see supplemental Table~\textcolor{blue}{S3} and Figs.~\textcolor{blue}{S3} and~\textcolor{blue}{S4}), the spin polarization remains uniformly oriented along the $z$-axis. This results in a persistent spin texture (PST)~\cite{LL-Tao2018}, distinct from the conventional Rashba and Dresselhaus types. Notably, the PST observed in monolayer Hf$_2$S is a truly full-space configuration, referred to as TFPST, and differs from previously reported FPSTs~\cite{JiPRBL2022,AbsorJPCM2022}, where spin degeneracy remains along certain momentum paths. In contrast, TFPST effectively suppresses spin dephasing and enables exceptionally long spin lifetimes, owing to a well-defined spin conservation axis across the entire BZ.

Given that only $s_{z}$ contributes to spin polarization, the spin state at energy level $E$ can be quantified as:
\begin{equation}
	\begin{aligned}
		s &= \sum_{n} s_{nk} = \sum_{n} \frac{A}{4\pi^2} \int \left\langle \psi_{n k} \right| s_{z} \left| \psi_{n k} \right\rangle \delta\left(E_{n k} - E\right) d k,
	\end{aligned}
\end{equation}
where $\left|\psi_{n k}\right\rangle$ is the eigenstate of band $n$ at $k$, and $A$ is the area of the 2D unit cell. The corresponding spin polarizability $p$ is defined as:
\begin{equation}
	\begin{aligned}
		p = \frac{s}{\sum_{n} \left|s_{nk}\right|},
	\end{aligned}
\end{equation}
where $p = 1$ or $-1$ corresponds to full spin-polarization.  As shown in the middle panel of Fig.~\ref{fig2}(c), $p$ in monolayer Hf$_{2}$S is clearly nonzero and can reach up to $\sim0.5$ under doping. In other type IV magnetic materials, such as Co$_{2}$P$_{4}$O$_{8}$, $p$ can even attain values of $\pm 1$ (see supplemental Fig. \textcolor{blue}{S2}), indicating the emergence of fully spin-polarized currents.

The symmetry $M_z$ enforces $\sigma_{xy}$ as the only nonzero AHC element. As shown in the right panel of Fig.~\ref{fig2}(c), although $\sigma_{xy}$ is nearly zero at the charge neutrality point, it increases to $-0.52~e^2/h$ at $0.55$ eV (P$_{1}$) and $0.86~e^2/h$ at $-0.42$ eV (P$_{2}$) under electron and hole doping, respectively. This clearly demonstrates that AHE can emerge in type IV magnetic materials even without nonrelativistic spin-splitting band structures.  At the energies of P$_{1}$ and P$_{2}$, $p$ is mostly zero, indicating that the AHE can be carried by spin-neutral currents with negligible spin polarization. Due to SOC, the magnetic moments of Hf atoms are slightly canted from the $y$-axis toward the $z$-axis, resulting in a tiny net magnetization of $\sim 10^{-4}~\mu_B$. Additional calculations with the Hf moments rigidly constrained along the $y$-axis confirm that the AHE does not originate from relativistic weak ferromagnetism.

\begin{figure}
	\centering
	\includegraphics[width=1\columnwidth]{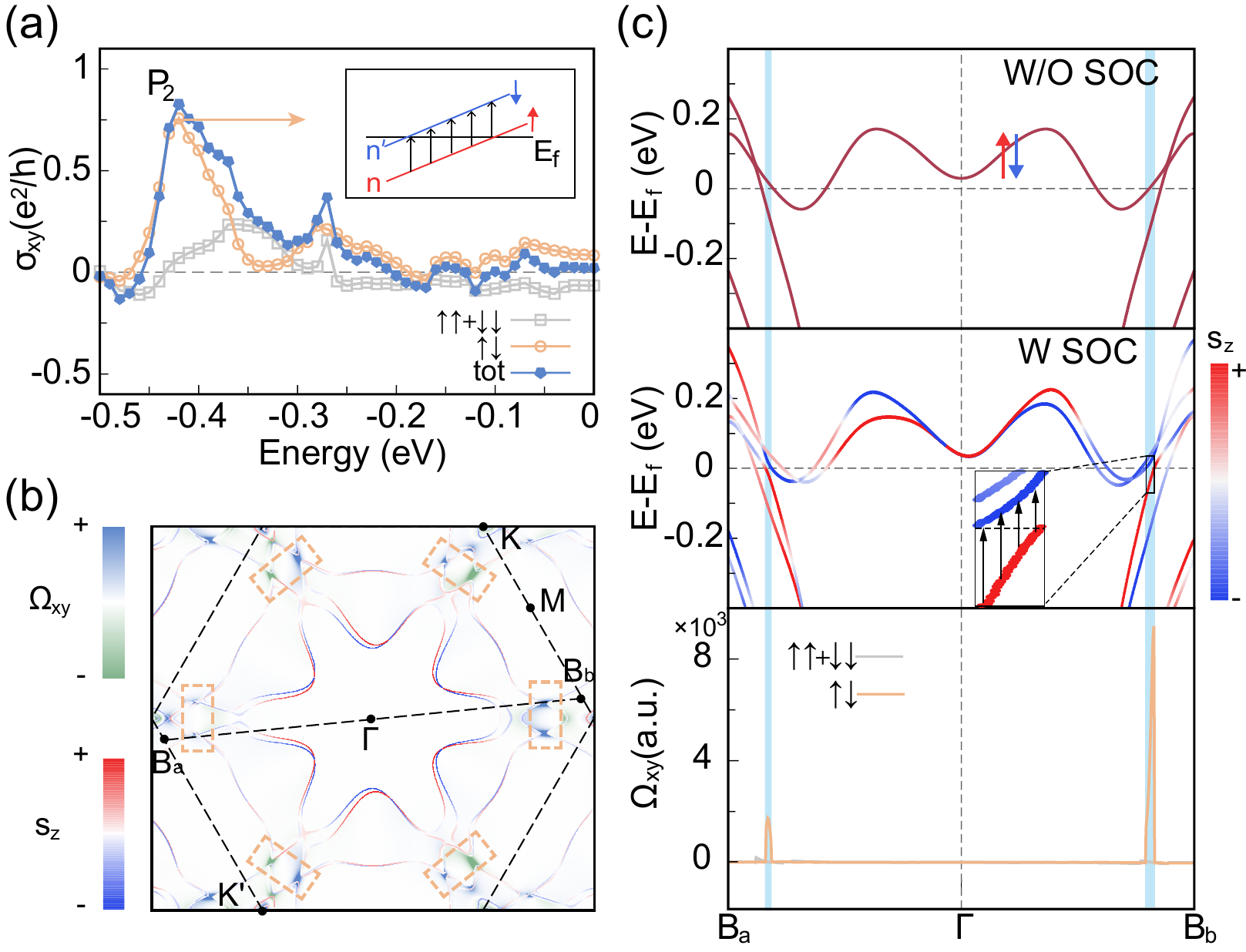}
	\caption{(a) Total anomalous Hall conductivity $\sigma_{xy}$ and its decomposition into spin-conserved ($\uparrow\uparrow + \downarrow\downarrow$) and spin-flip ($\uparrow\downarrow$) components as a function of the Fermi level. The inset illustrates a schematic diagram of ladder transitions. (b) $s_{z}$-resolved Fermi surface and Berry curvature $\Omega_{xy}$ at the energy of P$_{2}$. Spin-flip contributions are highlighted by orange dashed rectangles. (c) Nonrelativistic (top panel) and $s_{z}$-resolved relativistic (middle panel) band structures, along with Berry curvature distribution (bottom panel) at the energy of P$_{2}$. The inset in the middle panel shows a magnified view of the band structure, emphasizing the role of ladder transitions.}
	\label{fig3}
\end{figure}

To analyze the origin of the AHE, we plot the spin-decomposed $\sigma_{xy}$ under hole doping in Fig.~\ref{fig3}(a), where the spin-conserved (spin-flip) part corresponds to interband transitions between bands with the same (opposite) spin. The results show that near P$_{2}$, $\sigma_{xy}$ is mainly contributed by spin-flip transitions. This contrasts with ferromagnetic FePt~\cite{ZhangPRL2011} and altermagnetic RuO$_{2}$~\cite{ZhouPRL2024}, where spin-conserved processes typically dominate.  Figure~\ref{fig3}(b) shows the distribution of the Berry curvature together with the $s_{z}$-resolved Fermi surface at the energy of P$_{2}$. The Berry curvature hotspots are located in regions where a small band gap forms between two bands with opposite spin. Remarkably, this spin-flip contribution originates entirely from ladder transitions, arising from pairs of bands with parallel dispersion on either side of the Fermi level, as schematically illustrated in the inset of Fig.~\ref{fig3}(a).  To highlight this point, we select the $B_a$–$\Gamma$–$B_b$ path, where the calculated spin-split band structure and the associated Berry curvature peaks clearly demonstrate the role of ladder transitions in the spin-flip contribution [Fig.~\ref{fig3}(c)].  Finally, we calculate $\sigma_{xy}$ at a fixed Fermi energy as the N\'eel vector rotates within the $x$-$y$ plane (see supplemental Fig. \textcolor{blue}{S5}). In accordance with crystal symmetry, the $\sigma_{xy}$ curve exhibits a distinct threefold rotational pattern as the N\'eel vector changes its in-plane direction.  Moreover, $\sigma_{xy}$ inverts sign under the reversal of the Néel vector, enabling its use as an electrical probe of the Néel order.

In summary, we have uncovered a previously overlooked class of type IV 2D collinear magnets that lies beyond the conventional categorization of FMs, AFMs, and AMs. Characterized by nonrelativistic spin-degenerate bands and time-reversal symmetry-breaking responses such as the AHE, this phase bridges the gap between symmetry and transport phenomena in low-dimensional magnetism. Through systematic analysis of spin layer group and first-principles calculations, we identified realistic material candidates, among which monolayer Hf$_2$S serves as a representative example.  We demonstrate that, upon inclusion of SOC, spin-neutral currents can drive an AHE, accompanied by a TFPST protected by symmetry. These findings not only advance the theoretical understanding of 2D magnetic phases but also open new avenues for spintronic device design based on this novel class of 2D collinear magnetic materials.

This work is supported by the National Key R\&D Program of China (Grant No. 2022YFA1402600, No. 2022YFA1403800, and No. 2020YFA0308800), the National Natural Science Foundation of China (Grant No. 12274027, No. 12234003, and No. 12321004), and the Fundamental Research Funds for the Central Universities (Grant No. 2024CX06104).

\bibliography{mybib}

\end{document}